\documentclass[epj]{webofc}
\usepackage[varg]{txfonts}   
\usepackage{graphics}
\usepackage{graphicx}
\usepackage{epsfig}
\usepackage{amsmath}
\usepackage{amsfonts}
\usepackage[utf8]{inputenc}

\wocname{EPJ Web of Conferences}
\woctitle{CONF12}
\begin{document}
\selectlanguage{english}
\title{Recent results on the meson and baryon spectrum \\from lattice QCD}

\author{Daniel Mohler\inst{1,2}\fnsep\thanks{\email{mohler@kph.uni-mainz.de}}
}

\institute{Helmholtz-Institut Mainz, 55099 Mainz, Germany
\and
           Johannes Gutenberg Universit\"at Mainz, 55099 Mainz, Germany
}

\abstract{%
Recent lattice results on the meson and baryon spectrum with a focus on the
determination of hadronic resonance masses and widths using a combined basis
of single-hadron and hadron-hadron interpolating fields are reviewed. These
mostly exploratory calculations differ from traditional lattice QCD spectrum
calculations for states stable under QCD, where calculations with a  full
uncertainty estimate are already routinely performed. Progress and challenges
in these calculations are highlighted.}
\maketitle
\section{Introduction}
\label{intro}

In recent years, tremendous progress has been made in calculating QCD
observables from first principles using Lattice QCD. In particular, it is now possible to
perform calculations at physical light-quark masses and with multiple lattice
spacings and volumes. Improved  actions have been developed for simulations
with both light (u,d,s) and heavy (c,b) quarks, allowing for a reliable
extraction of simple observables for mesons made from these quarks.

\begin{figure}[tbh]
\centering
\sidecaption
\includegraphics[clip,width=0.5\textwidth]{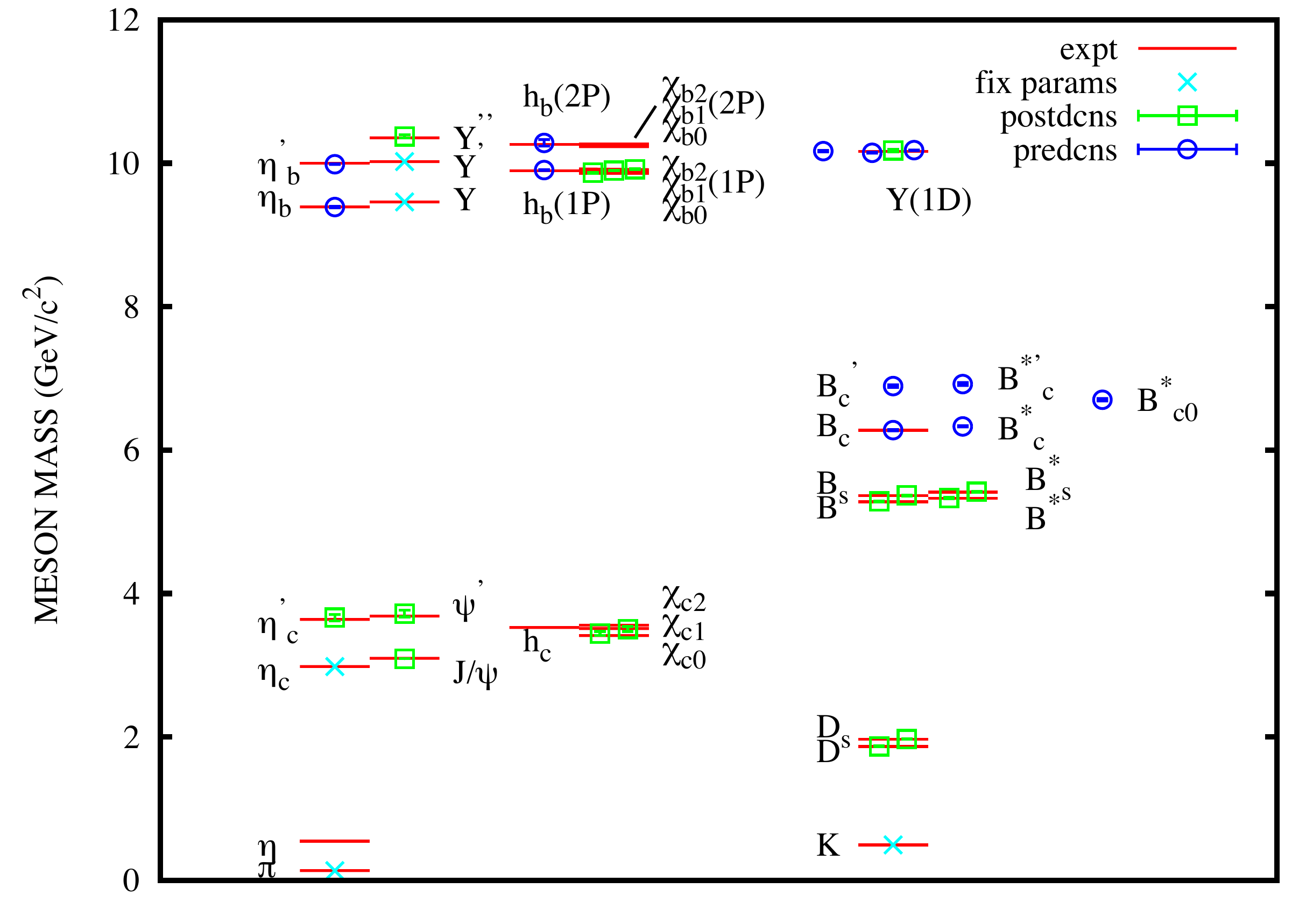}
\caption{The spectrum of mesons comparing HPQCD lattice QCD
  results to experiment. Crosses are quantities used to fix parameters in the
  action, while the other results were either postdictions or
  predictions. Figure from \cite{Dowdall:2012ab}.}
\label{hpqcd}
\end{figure}

Figure \ref{hpqcd} (from \cite{Dowdall:2012ab}) shows results for the lattice
spectrum of states stable or nearly stable under the strong interaction
compared to experiment. After some of the experimental masses are used as input to tune
the mass parameters and coupling in the action (cyan crosses in the figure),
parameter free theory postdictions and predictions were obtained.

\begin{figure}[tbh]
\begin{center}
\sidecaption
\includegraphics[clip,width=0.5\textwidth]{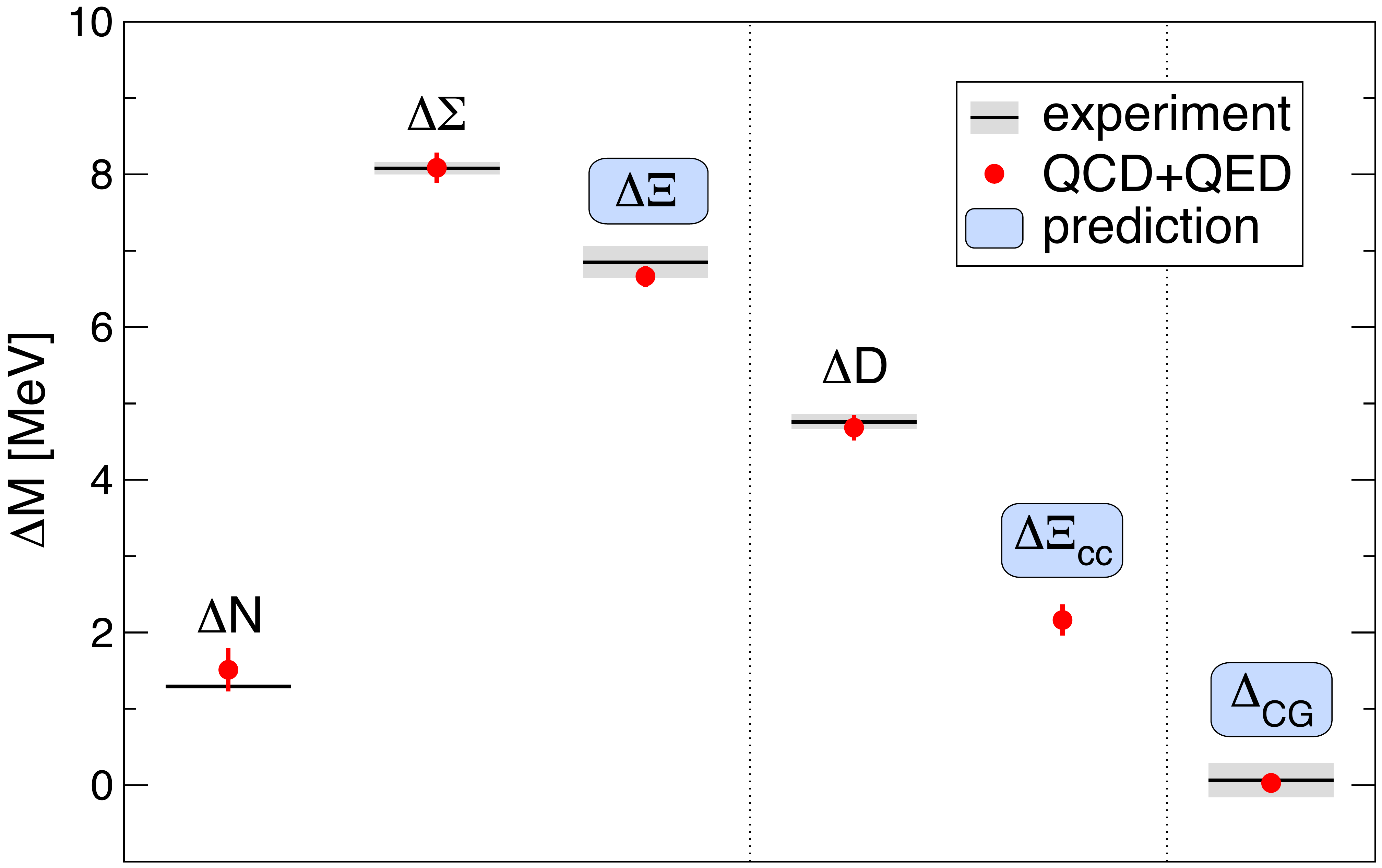}
\caption{\label{bmw} Mass splittings in channels stable under the strong and
  electromagnetic interactions. The red dots and uncertainties are the lattice data
  and the lines and shaded bands are the experimental results. Figure from \cite{Borsanyi:2014jba}.}
\end{center}
\end{figure}

Beyond pure QCD, recent progress has allowed the dynamical simulation of both
QCD and QED on the lattice. Figure \ref{bmw} (from \cite{Borsanyi:2014jba}) shows the isospin splitting of
various ground state hadrons, resulting from both the
difference of up and down quark masses and from QED. The resulting mass
splittings agree perfectly with those measured in experiment, while splittings
not yet observed to date in experiment can also be calculated.

In these proceedings we address more challenging observables, taking a look at
hadrons close to multiparticle thresholds and hadronic resonances. While the
masses of stable states well separated from multi-particle thresholds are
readily extracted in modern lattice calculations, extracting the spectrum of
hadronic excitations is a challenge. In Section \ref{luescher} readers will be
pointed to literature relevant to understand theoretical aspects of extracting
scattering observables from Lattice QCD. In the remainder of Section \ref{res}
the current state of the art with regard to both meson-meson and meson-baryon
scattering are briefly reviewed by appealing to specific examples from the
literature. In Section \ref{mesons} examples for meson-meson scattering
studies will be presented, while \ref{lesson} highlights some of the findings
from these studies. In Section \ref{baryons} progress in the simulation of
meson-baryon scattering is discussed. Section \ref{outlook} provides a brief outlook.

\section{Spectroscopy of resonances and close-to-threshold states}
\label{res}
\subsection{L\"uscher's finite volume method}
\label{luescher}

\begin{figure}[tbh]
\begin{center}
\sidecaption
\includegraphics[clip,height=4.5cm]{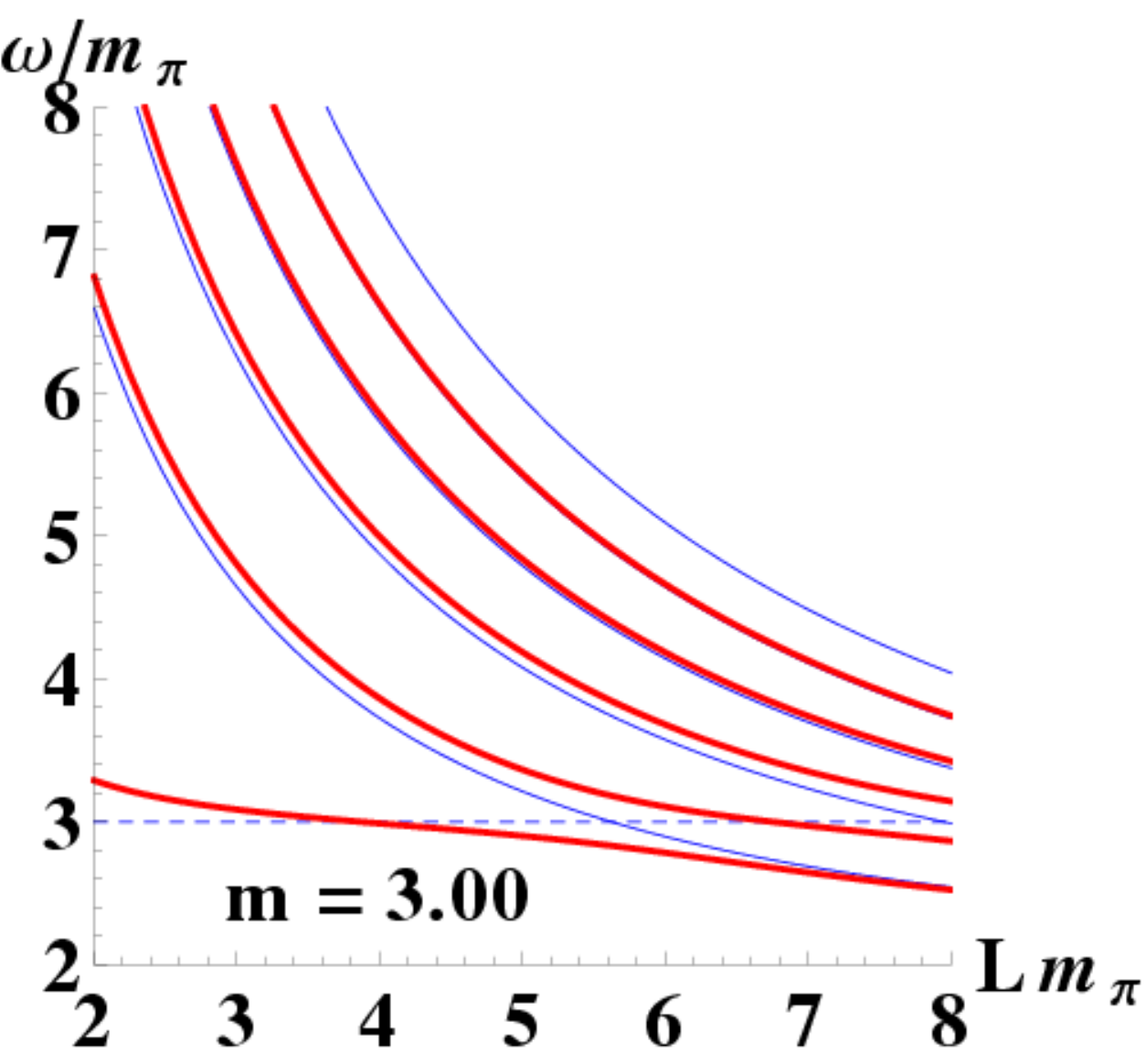}
\caption{\label{luscher} Energies for the scattering of two bosons in a channel with a resonance at
three times the boson mass as a function of the box size. The blue lines
represent the non-interacting spectrum, while the red lines displaying a
characteristic avoided level crossing pattern show the expectation in presence
of the interaction.}
\end{center}
\end{figure}

From Euclidean space correlation functions, scattering observables are not
accessible directly. In a series of seminal papers by L\"uscher \cite{Luscher:1986pf,Luscher:1990ux,Luscher:1991cf} it
was pointed out that information about the continuum scattering amplitude of
elastic scattering can be inferred from the finite volume dependence of the
energy levels, which appear shifted from the free energy levels due to the interaction.
These energy shifts are illustrated in Figure \ref{luscher}, where the
energies for the scattering of two bosons in a channel with a resonance at
three times the boson mass are drawn as a function of the box size. 

The relations first provided by L\"uscher have been greatly generalized over the
years, now including relations for any number of (coupled) two particle
channels, for $2\leftrightarrow 1$ and $2\leftrightarrow2$ transitions (such
as $\pi\pi\rightarrow\pi\gamma^*$), for particles with and without spin, for moving frames,
\textit{etc.}. For recent reviews of the formalism including the relevant
references please refer to \cite{Briceno:2014pka,Hansen:2015azg}. In practical
applications it is crucial to be able to obtain enough lattice energy levels
in the region of interest. To this end both multiple volumes and multiple
momentum frames are used.

\subsection{Meson-meson scattering and resonances}
\label{mesons}

\subsubsection{The $\rho$ meson}
\label{rho}
The simplest QCD resonance is the $\rho$ meson with quantum numbers
$J^{PC}=1^{--}$, seen in isospin-1 $\pi\pi$ scattering. It decays nearly 100\%
of the time into two pions \cite{Olive:2016xmw} and there seem to be no significant inelastic
contributions from the open four pion threshold. For these reasons the $\rho$
meson is the ideal benchmark for lattice studies of elastic scattering
employing the L\"uscher formalism. After a pioneering study by the CP-PACS
collaboration \cite{Aoki:2007rd}, multiple groups provided proof of principle
calculations demonstrating the feasibility of lattice resonance studies \cite{Feng:2010es,Lang:2011mn,Aoki:2011yj}. In the meantime a number of further studies exist \cite{Pelissier:2012pi,Dudek:2012xn,Wilson:2015dqa,Bali:2015gji,Bulava:2016mks,Guo:2016zos}.

\begin{figure}[tbh]
\begin{center}
\sidecaption
\includegraphics[clip,width=0.6\textwidth]{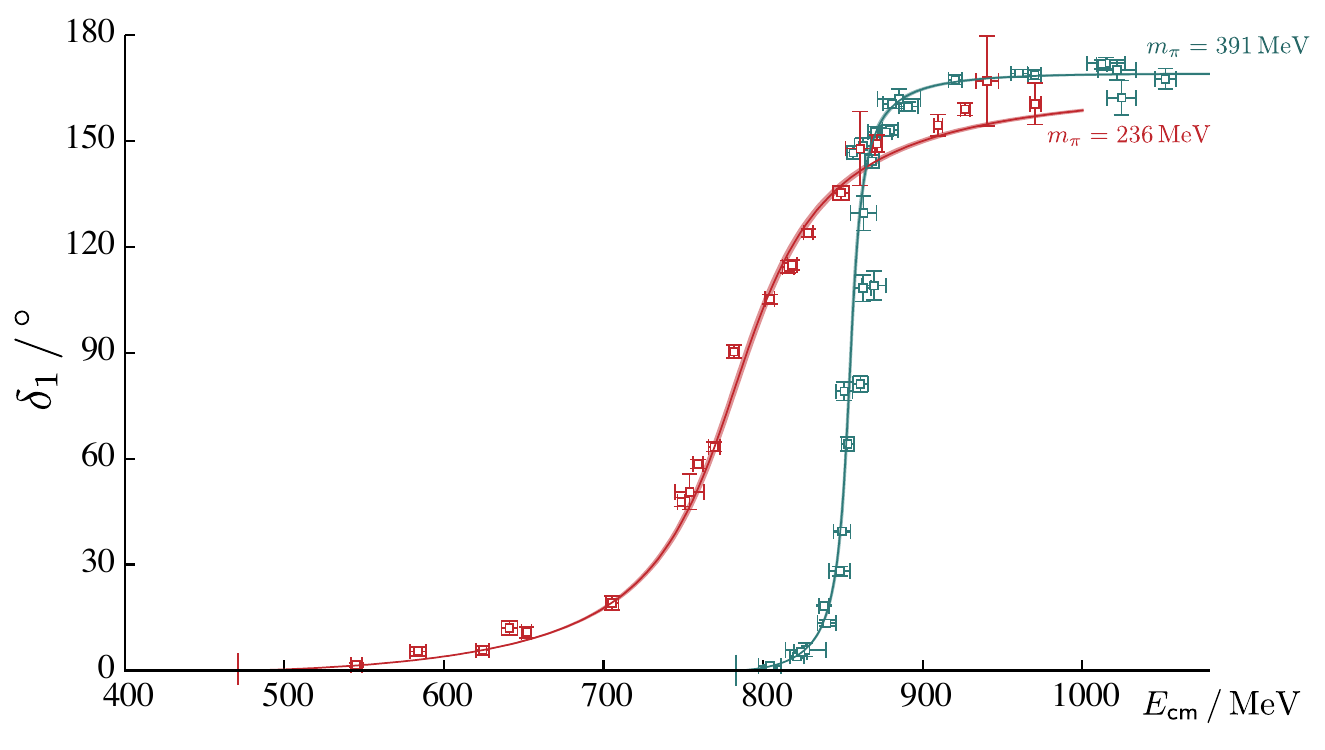}
\caption{\label{rho_jlab} Elastic P-wave $\pi\pi$ scattering phase shifts with
isospin 1 for two different pion masses. The blue data points are for $M_\pi=391$~MeV
and the red data points for $M_\pi=236$~MeV. Figure from \cite{Wilson:2015dqa}.}
\end{center}
\end{figure}

Figure \ref{rho_jlab} shows an example of a state-of-the-art calculation by
the Hadron Spectrum Collaboration \cite{Wilson:2015dqa}. The figure shows results for the p-wave
phase shift $\delta_1$ of elastic $\pi\pi$ scattering for two unphysically large
pion masses as a function of the center of mass energy $E_{cm}$. Using a
Breit-Wigner parameterization, the resonance mass and coupling $g_{\rho\pi\pi}$
can be extracted. The large number of lattice data points results from using
all lattice irreps for multiple center-of-mass momenta, and multiple lattice volumes.

\begin{figure}[tbh]
\begin{center}
\sidecaption
\includegraphics[clip,width=0.5\textwidth]{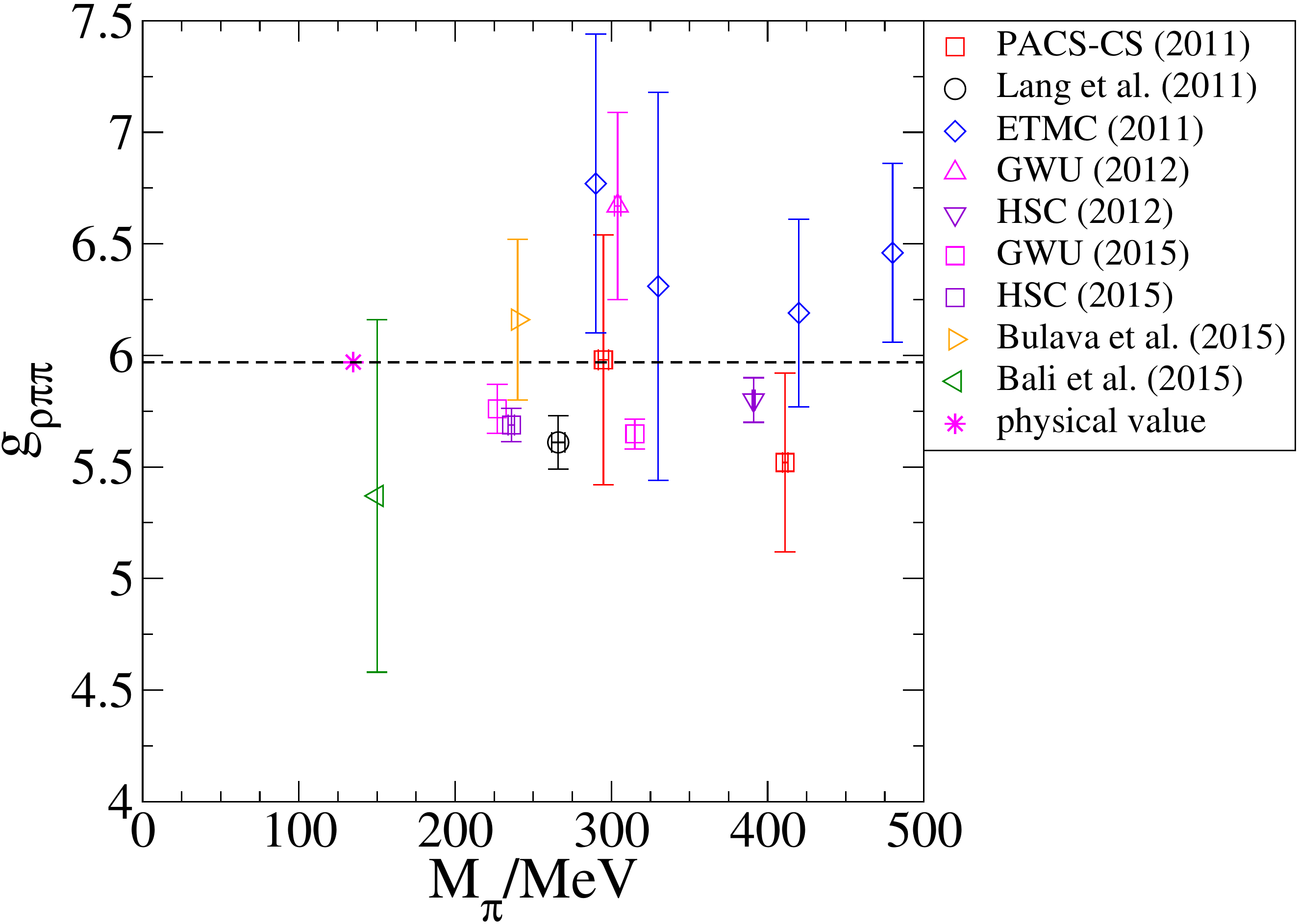}
\caption{\label{grhopipi} Collection of results from various lattice
  determination of the coupling $g_{\rho\pi\pi}$. For a list of references
  please refer to the text.}
\end{center}
\end{figure}

Figure \ref{grhopipi} shows the current world data for $g_{\rho\pi\pi}$ from
the lattice determinations \cite{Feng:2010es,Lang:2011mn,Aoki:2011yj,Pelissier:2012pi,Dudek:2012xn,Wilson:2015dqa,Bali:2015gji,Bulava:2016mks,Guo:2016zos}. While there is in
general a good agreement among the lattice data, the uncertainties in the plot are (in most cases)
statistical only and there is no calculation demonstrating a full control of
systematic uncertainties. 

\begin{figure}[tbh]
\begin{center}
\sidecaption
\includegraphics[clip,width=0.5\textwidth]{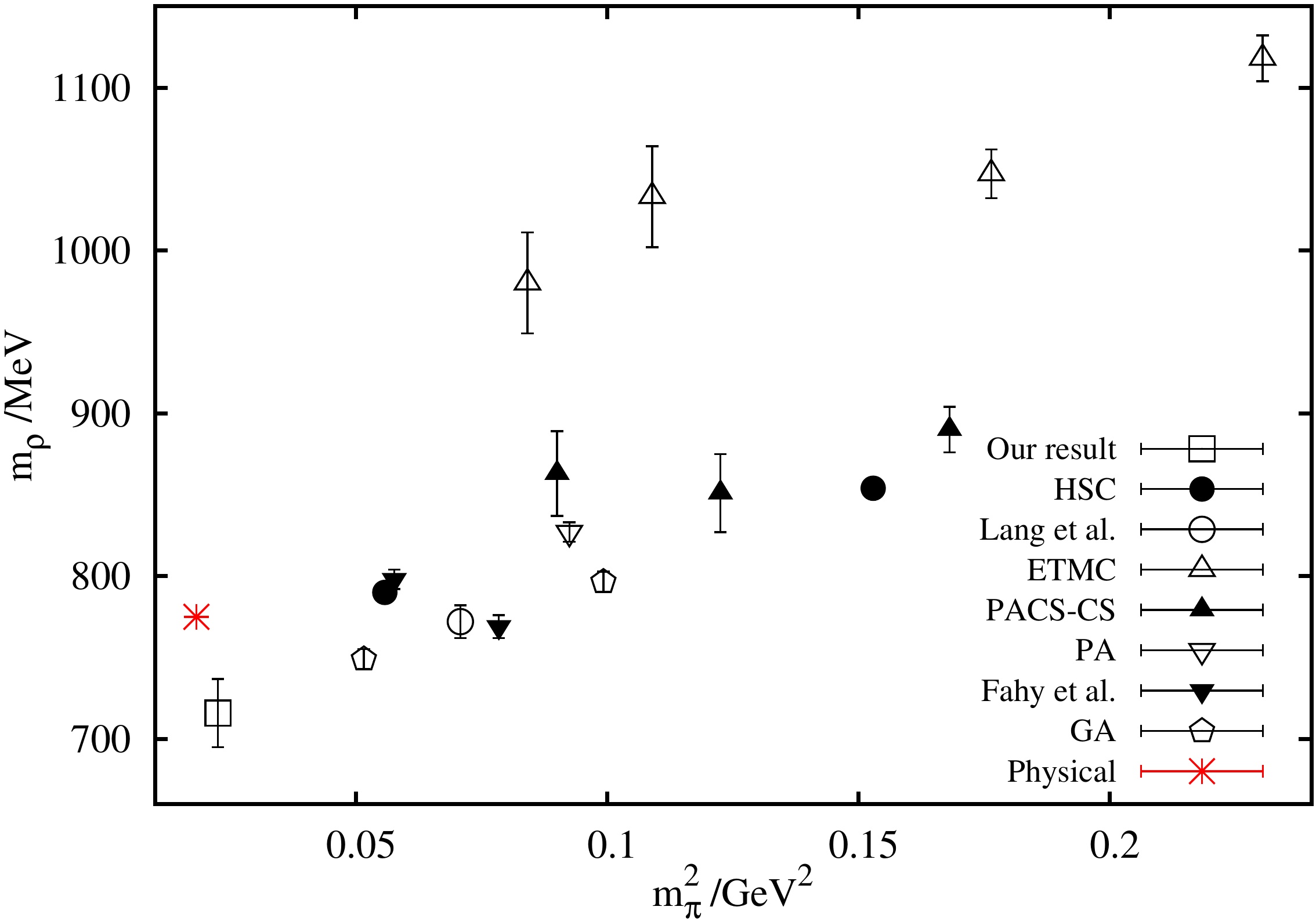}
\caption{\label{mrho} Collection of results from various lattice
  determination of the $\rho$ meson resonance mass taken from
  \cite{Bali:2015gji}. Notice that deviations are expected at non-physical
  pion masses as different groups use different ways to determine the physical quark masses and to set
the scale in their lattice calculations.}
\end{center}
\end{figure}

A similar plot for the extracted resonance masses is shown in Figure
\ref{mrho}, taken from \cite{Bali:2015gji}. It illustrates that the
neglected systematics are likely important, as the resonance masses from
different simulations tend to deviate quite a bit from each other at
unphysical pion masses. Notice however that some deviations are expected as different
groups use different ways to determine the physical quark masses and to set
the scale in their lattice calculations. As a dimensionful quantity, the
resonance mass therefore should only agree for physical parameters and in the continuum
limit. Demonstrating this agreement will be a task for future lattice simulations.

\begin{figure}[tbh]
\begin{center}
\includegraphics[clip, height=3.0cm]{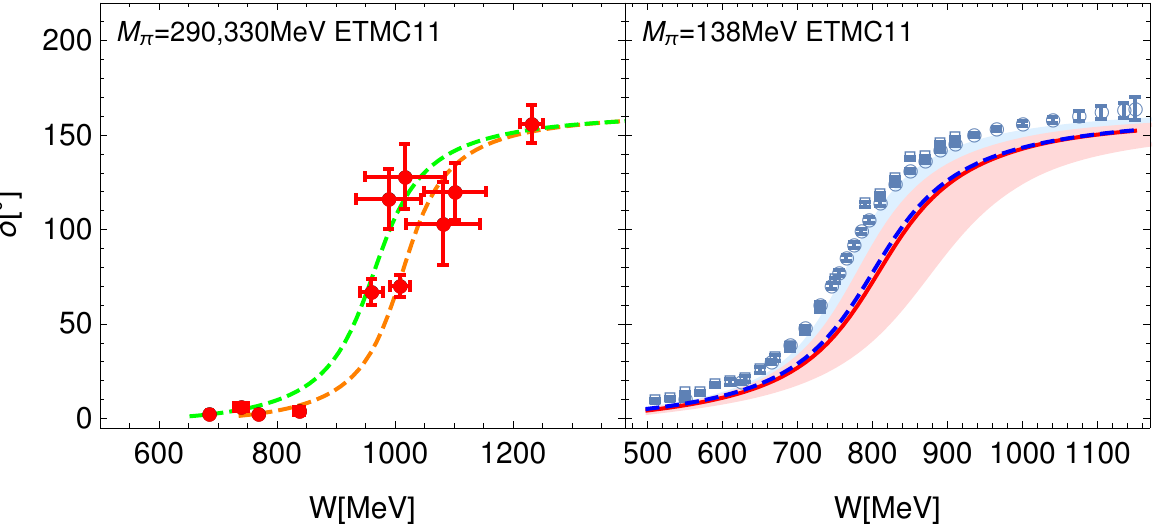}
\includegraphics[clip, height=3.0cm]{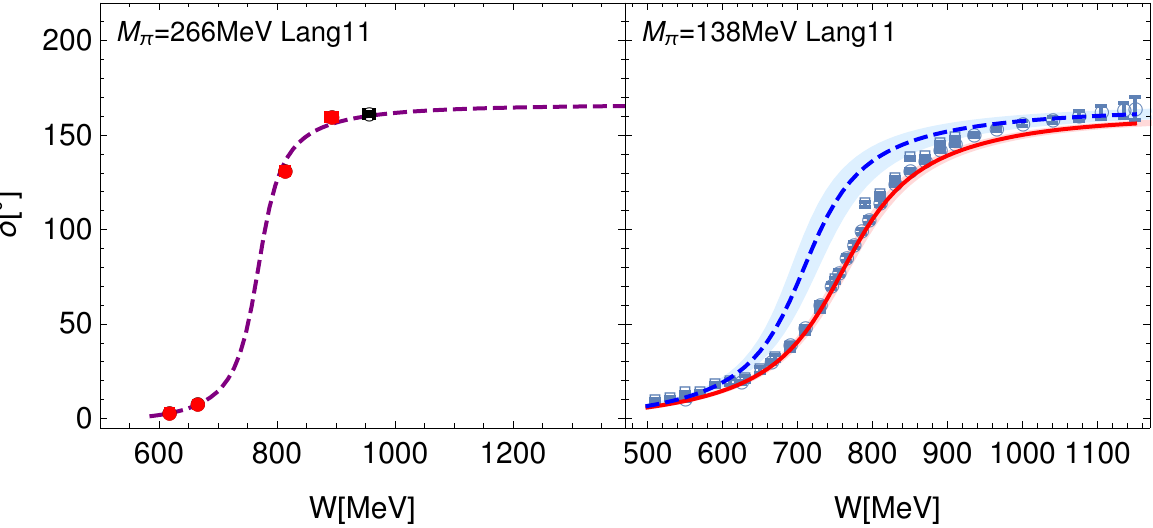}\\
\includegraphics[clip, height=3.0cm]{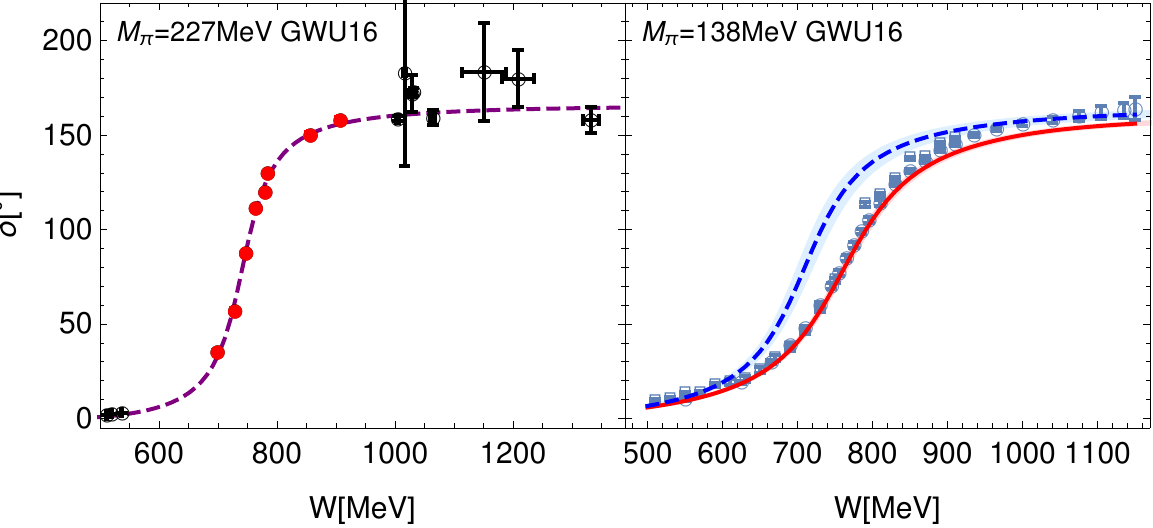}
\includegraphics[clip, height=3.0cm]{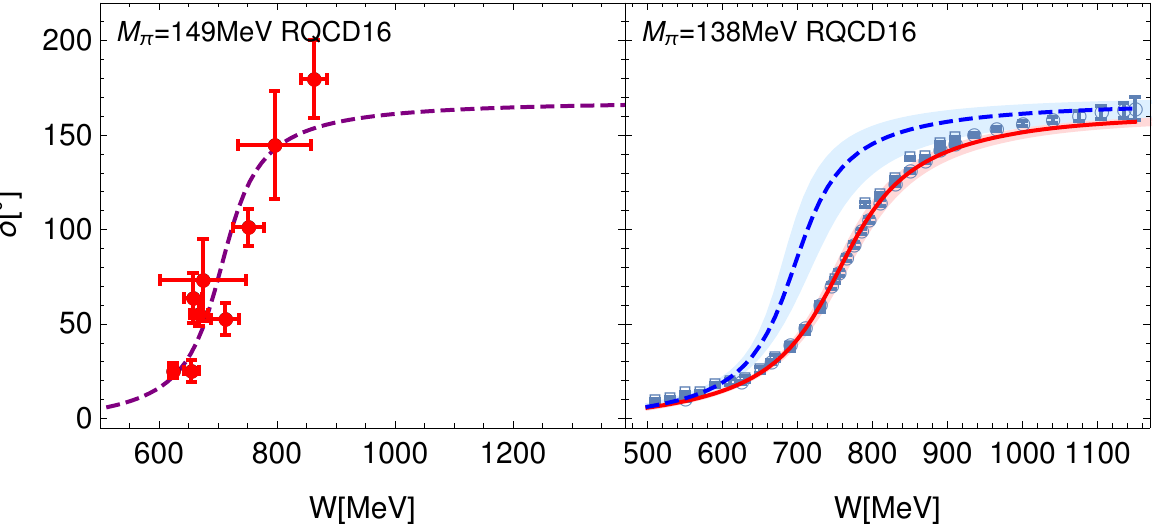}
\caption{\label{rho_2flavor} The left side in each plot shows lattice results
  from 2-flavor simulations of the $\rho$ meson from a subset of results
  analyzed in \cite{Hu:2016shf}. The right side in each plot shows the
  chirally extrapolated 2-flavor data (blue dashed curves and blue error
  bands) and the results after including the $\bar{K}K$ channel (red curves and
  bands) compared to the experimental data (blue circles and squares). Plot
  from \cite{Hu:2016shf}.}
\end{center}
\end{figure}

With the growing dataset for the $\rho$-resonance the question naturally
arises if there are additional insights gained from taking a more detailed
look at the current set of results. In \cite{Hu:2016shf} the authors used Unitarized
Chiral Perturbation Theory (Unitarized $\chi$PT) to analyze the lattice
scattering data with 2 flavors of light (up and down) dynamical quarks. Fitting to
experiment data in addition, the Unitarized $\chi$PT calculation allows to
``switch on'' the strange quarks. As their final result the authors claim that
the low resonance masses observed in 2 flavor calculations (compared to the
physical rho and the corresponding 2+1 flavor calculations) are a result of
the missing $\bar{K}K$ channel. If confirmed this constitutes a striking effect
from the (partial) quenching of the strange quark. Future lattice calculations
with both 2 and 2+1 flavors of dynamical quarks will shed light on this issue.

\begin{figure}[tbh]
\begin{center}
\sidecaption
\includegraphics[clip,width=0.6\textwidth]{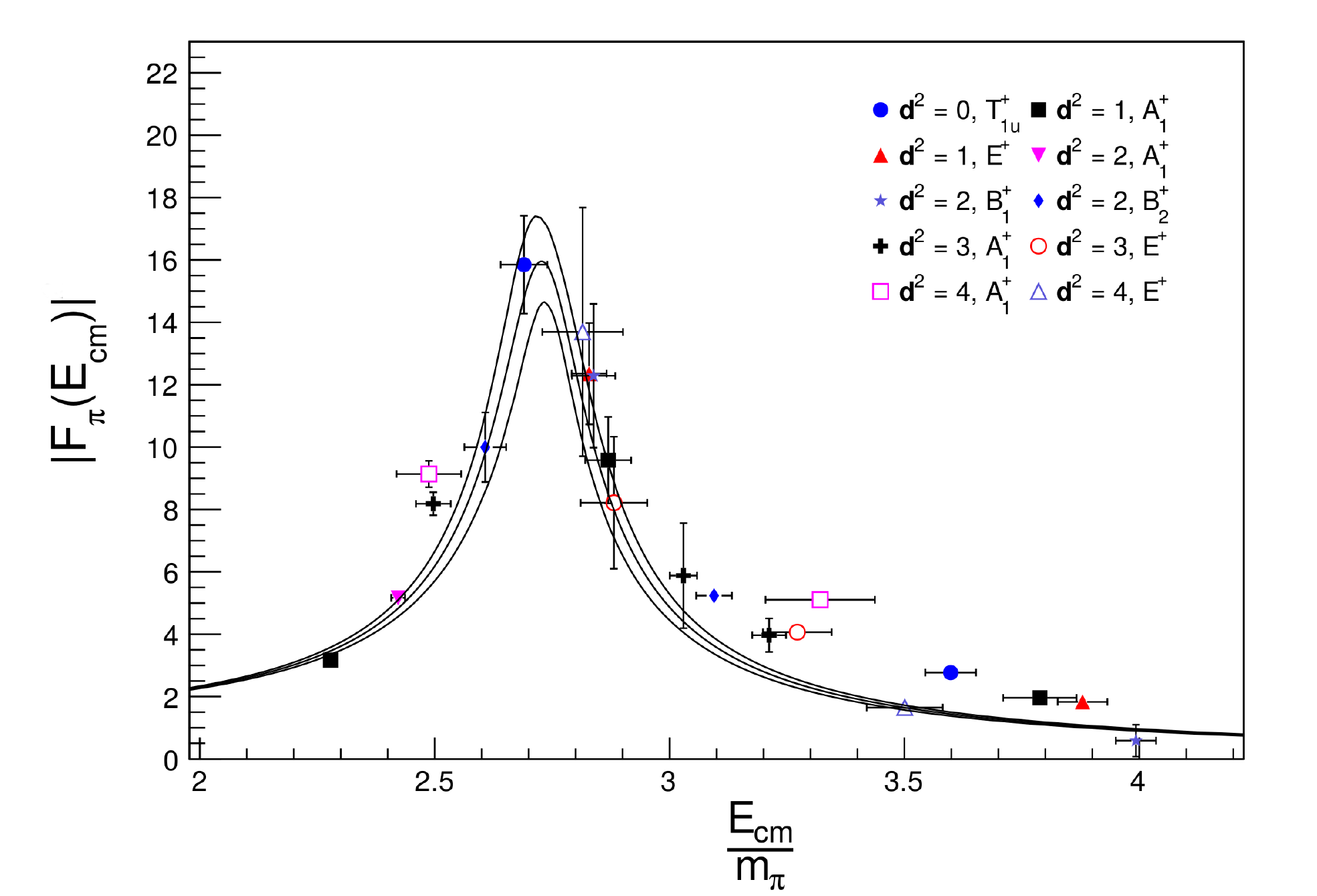}
\caption{\label{fpi_dublin} $\lvert F_\pi(\sqrt(s))\rvert$ as extracted from a
  lattice QCD calculation at $m_\pi=240$~MeV in \cite{Bulava:2015qjz}. The
  curve corresponds to a Gounaris-Sakurai parameterization with $m_\rho$ and
  $g_{\rho\pi\pi}$ determined from the lattice data.}
\end{center}
\end{figure}

Beyond the resonance mass and couplings, a lattice determination of the phase
shift $\delta_1$ and its derivative also allows for the determination of the
pion timelike form factor in the elastic region $2m_\pi\le\sqrt{s}\le 4m_\pi$
\cite{Meyer:2011um}. In this region the form factor  $F_\pi(E)$ is given by
\begin{align}
R(s)&=\frac{1}{4}\left(1-\frac{4m_\pi^2}{s}\right)^\frac{3}{2}\lvert
F_\pi(\sqrt{s})\rvert^2=\frac{\sigma(e^+e^-\rightarrow hadrons)}{4\pi\alpha(s)^2/(3s)}\;.
\end{align}
It is of phenomenological importance for lattice determinations of the hadronic vacuum
polarization (HVP) contribution to $(g-2)_\mu$ \cite{Bernecker:2011gh}. Recently first
lattice calculations of the form factor have been performed in
\cite{Feng:2014gba,Bulava:2015qjz}. Figure \ref{fpi_dublin} shows results at unphysical
parameters from \cite{Bulava:2015qjz}, compared
to a Gounaris-Sakurai parameterization (not a fit). The displayed results are
from a single 2+1 flavor ensemble with $m_\pi\approx 280$. These recent calculations demonstrate that
the timelike pion form factor can be extracted with a reasonable
precision.

\subsubsection{$D_{s0}^*(2317)$ and $D_{s1}(2460)$ and their b-quark cousins}
\label{dmesons}

In the spectrum of positive parity $D_s$ mesons, the $J^P=0^+$
$D_{s0}^*(2317)$ and $1^+$ $D_{s1}(2460)$ have masses and properties not expected
from potential models for $\bar{q}q$ mesons. A particularly peculiar fact is
that their mass is essentially degenerate with the corresponding $D$ mesons,
even though the strange quark is much heavier than the light up and down
quarks. This fact lead to speculations that these states have an exotic
structure, such as a tetraquark or molecular structure. 

\begin{figure}[tbh]
\begin{center}
\sidecaption
\includegraphics[clip,width=0.6\textwidth]{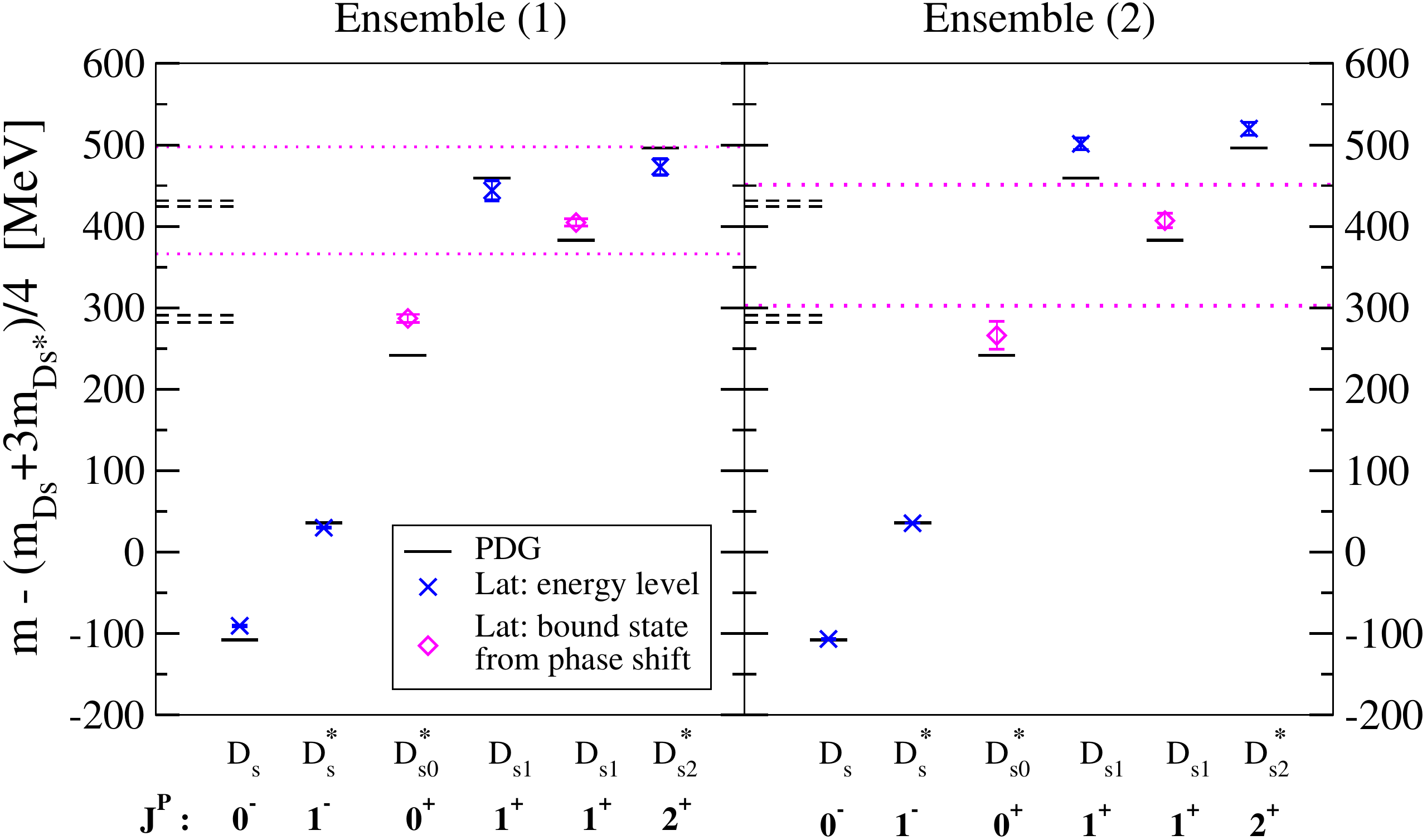}
\caption{\label{ds_spectrum} Low-lying $D_s$ meson spectrum from \cite{Mohler:2013rwa,Lang:2014yfa}
  presented with respect to spin-averaged mass
  $\tfrac{1}{4}(m_{D_s}+3m_{D_s^*})$. The diamonds and crosses display our
  lattice results, while black full lines correspond to experiment.  The
  magenta diamonds correspond to the pole position in the $T-$matrix. Masses
  extracted as energy levels in a finite box are displayed as blue
  crosses. Dotted (dashed) lines correspond to $DK$ and $D^*K$ lattice
  (physical) thresholds.}
\end{center}
\end{figure}

While traditional lattice
studies using just quark-antiquark interpolating fields tend to get too large
or badly determined masses for those states, a more recent study \cite{Mohler:2013rwa,Lang:2014yfa} using
both quark-antiquark and $D^{(*)}K$ structures leads to a spectrum in
qualitative agreement with experiment, highlighting the role of the nearby
thresholds. The results from this study are shown in Figure \ref{ds_spectrum}.

\begin{figure}[tbh]
\begin{center}
\includegraphics[clip,height=3.4cm]{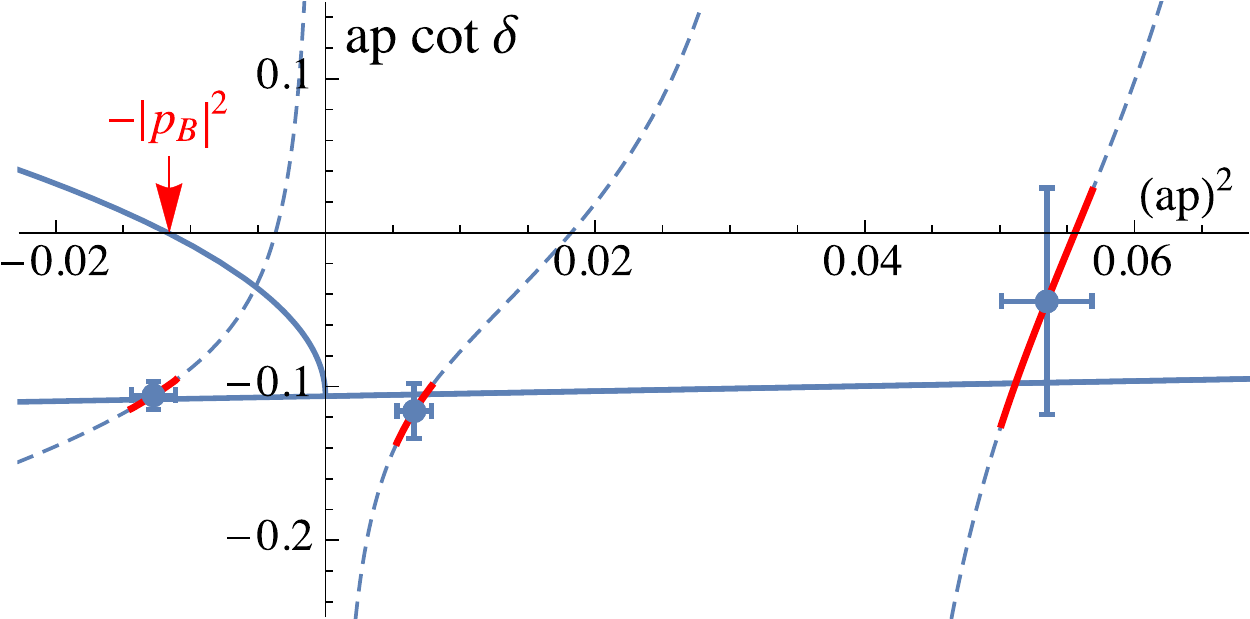}
\includegraphics[clip,height=3.4cm]{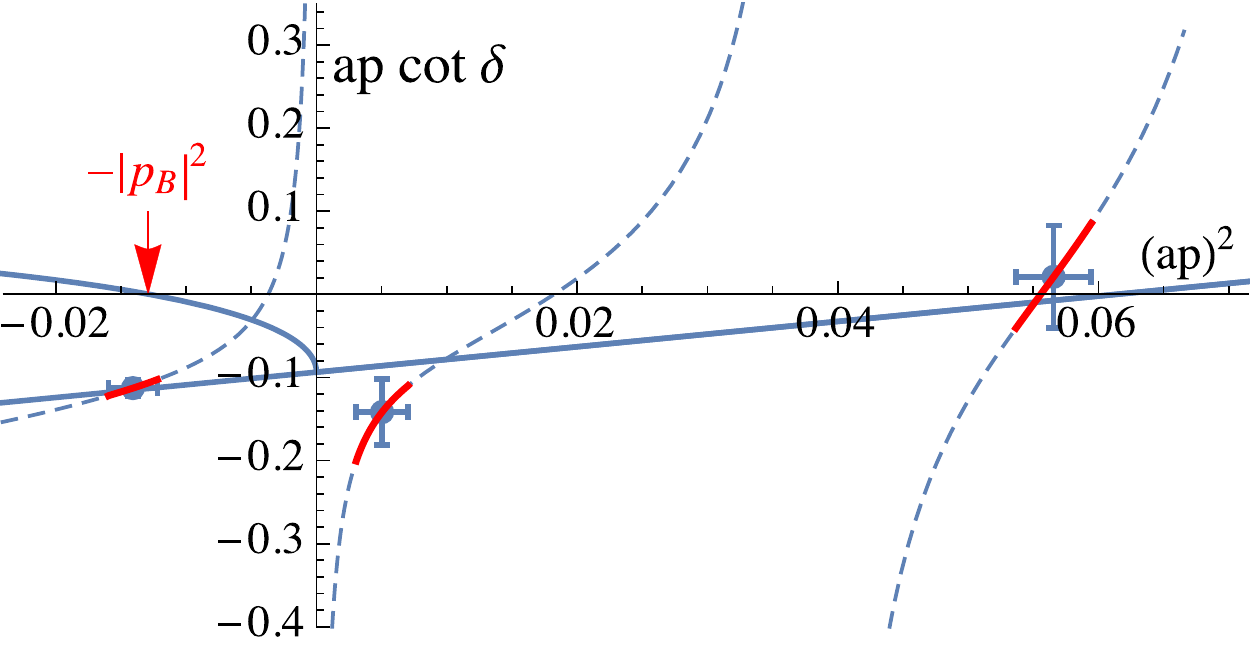}
\caption{\label{bs0bs1} Plots of $ap \cot\delta(p)$ vs. $(ap)^2$ for
  $B^{(*)}K$ scattering in $s$-wave from \cite{Lang:2015hza}. Circles are values from our
  simulation; red lines indicate the error band following the L\"uscher curves
  (dashed lines). The solid line gives the effective range fit to the
  points. The values for $-\lvert p_B\rvert^2$ corresponding to the binding
  energy in infinite volume are indicated by the arrows. Displayed uncertainties are statistical only.}
\end{center}
\end{figure}

\begin{figure}[tbh]
\begin{center}
\sidecaption
\includegraphics[clip,height=5.0cm]{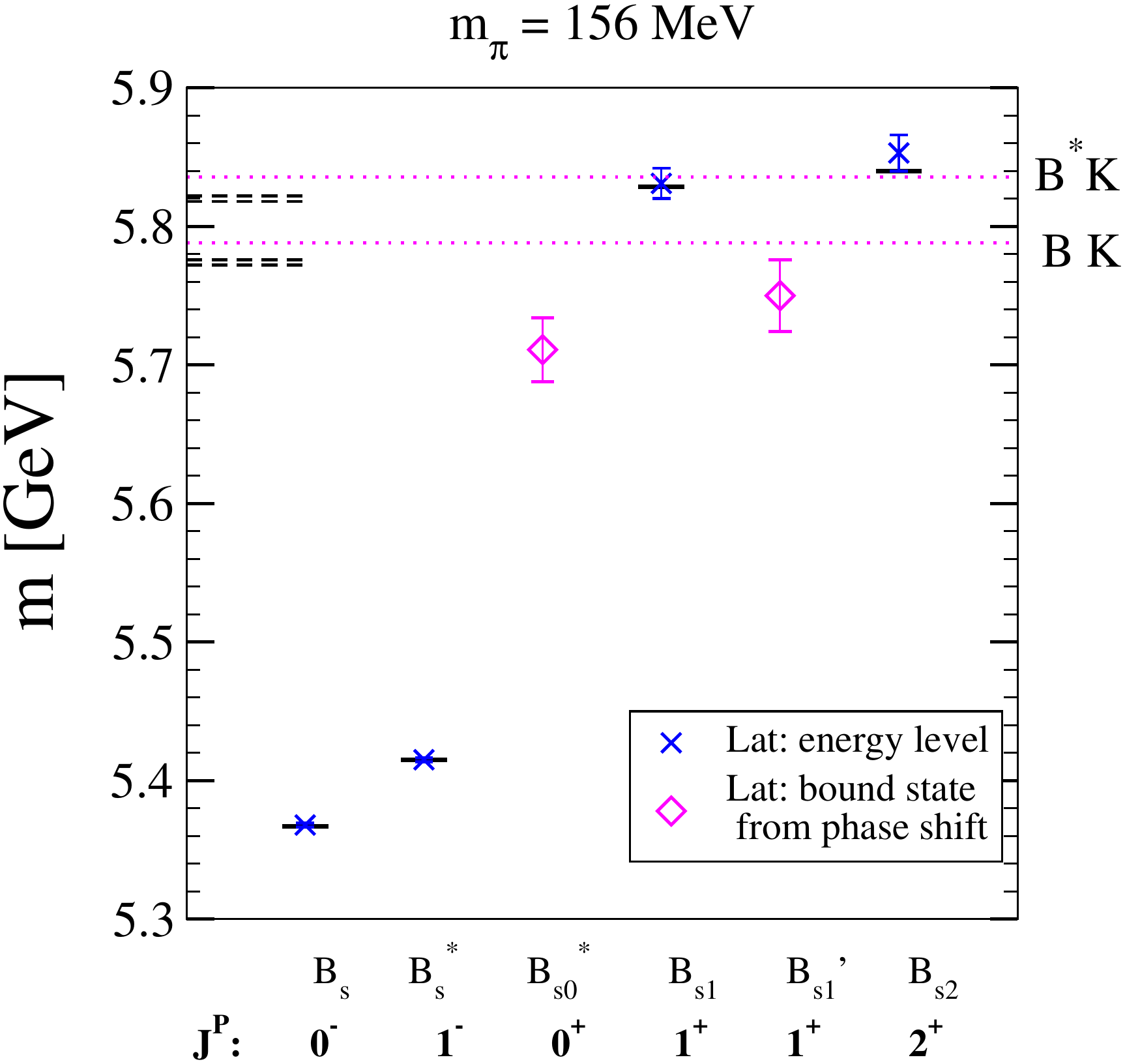}
\caption{\label{bs_spectrum} Spectrum of s-wave and p-wave $B_s$ states from \cite{Lang:2015hza}. The
  blue states are naive energy levels, while the bound state energy of the
  states in magenta results from an effective range approximation of the phase
shift data close to threshold. The black lines are the energy levels from the
PDG \cite{Olive:2016xmw}. The error bars on the blue states are
statistical only, while the errors on the magenta states show the full
(statistical plus systematic) uncertainties. }
\end{center}
\end{figure}

While the LHCb experiment should be able to observe them, the corresponding
$0^+$ and $1^+$ states in the $B_s$ spectrum are not yet observed in
experiment. Using the finite volume formalism described in Section
\ref{luescher} and performing an effective range approximation close to
threshold, a recent study \cite{Lang:2015hza} observed two bound states below the $BK$
and $B^*K$ thresholds (respectively). The lattice data along with the
effective range parameterization is plotted in Figure \ref{bs0bs1}. Figure
\ref{bs_spectrum} shows the resulting spectrum for the 1S and 1P
states. Notice that the known $1^+$ and $2^+$ mesons above threshold \cite{Olive:2016xmw}
are well reproduced. For more details please refer to \cite{Lang:2015hza}.

\subsubsection{Coupled channel scattering and light scalar mesons}
\label{scalars}

Using 3 lattice volumes and several moving frames, the Hadron Spectrum
Collaboration has started perform the first coupled-channel analyses of
meson-meson systems in a finite volume. So far they have investigated coupled
$\pi K$--$\eta K$ scattering \cite{Dudek:2014qha,Wilson:2014cna}, $\pi\eta$--$K\bar{K}$ scattering \cite{Dudek:2016cru}, and
extended their earlier $\rho$ meson study by also considering coupled channel
$\pi\pi$--$K\bar{K}$ scattering \cite{Wilson:2015dqa}.

\begin{figure}[tbh]
\begin{center}
\includegraphics[clip,width=1.0\textwidth]{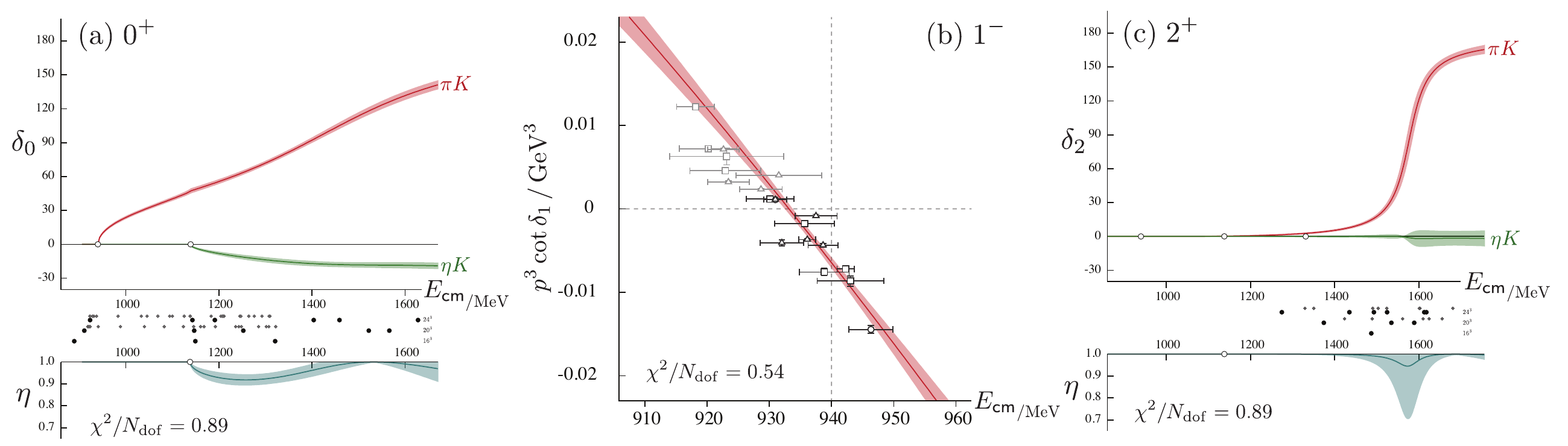}\\
\caption{\label{pik_etak} Results for $\pi K$ and $\eta K$ phase shifts with
  quantum numbers $J^P=0^+$ (a), $1^+$ (b), and $2^+$ (c). For the coupled
  channel scattering with quantum numbers $0^+$ and $2^+$ the inelasticity
  $\eta$ is also plotted. Plot from \cite{Dudek:2014qha}}
\end{center}
\end{figure}

\begin{figure}[tbh]
\begin{center}
\sidecaption
\includegraphics[clip,width=0.6\textwidth]{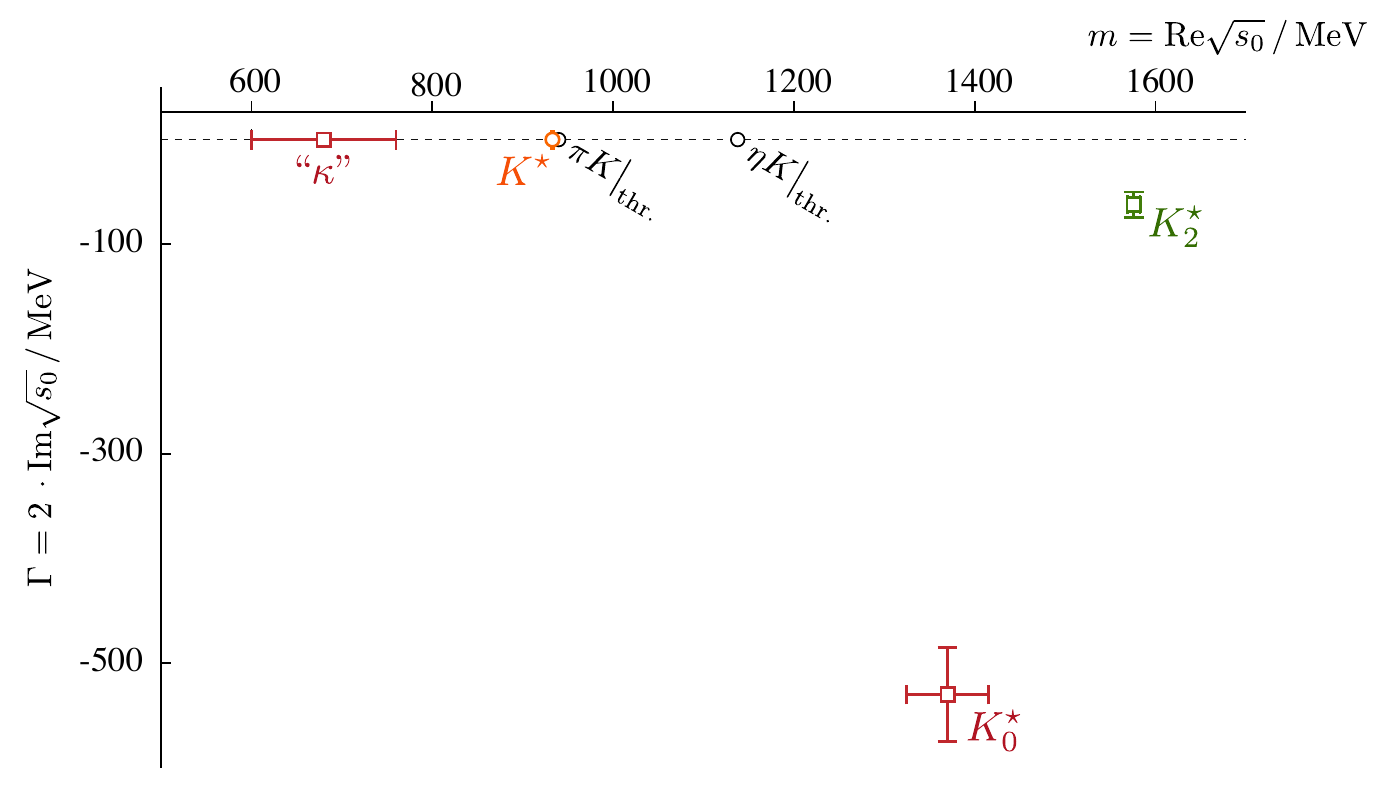}
\caption{\label{poles} Pole positions in the complex plain for states with
  quantum numbers $0^+$ (red) $1^-$ (orange) and $2^+$ (green). For a full
  description please refer to \cite{Dudek:2014qha,Wilson:2014cna}.}
\end{center}
\end{figure}

Figure \ref{pik_etak} shows their results for coupled channel $\pi K$--$\eta
K$ scattering with $J^P=0^+$, $1^-$ and $2^+$. The low-lying physical
resonances \cite{Olive:2016xmw} in these channels are the $K_0^*(800)$ (also called $\kappa$) and the
$K_0^*(1430)$ for $J^P=0^+$, the $K^*(892)$ for $J^P=1^-$, and the
$K^*_2(1430)$ for $J^P=2^+$. The lattice results at an larger than physical
pion mass of $391$~MeV qualitatively agree with the physical spectrum. In the
$0^+$ channel shown in the left panel of Figure \ref{pik_etak}, a virtual
bound state related to the physical $\kappa$ resonance is found along a much
heavier and very broad $K_0^*$ resonance. The broad resonance is
qualitatively compatible with the $K_0^*(1430)$. At such a large pion mass the
$K^*(892)$ is not a resonance but a close-to-threshold bound state, and the
corresponding coupling is compatible with the coupling extracted from
experiment. The energy levels from which this bound state can be determined
are depicted in the mid pane of the figure. An earlier single-channel lattice
calculation \cite{Lang:2012sv,Prelovsek:2013ela} at a lighter pion mass of $266$~MeV already observed a
very narrow resonance \cite{Prelovsek:2013ela}, also compatible with the $K^*(892)$. The
right panel of Figure \ref{pik_etak} shows the lattice data for the $J^P=2^+$
channel, where a narrow resonance compatible with the $K^*_2(1430)$ is
observed. Above the $\eta K$ threshold the inelasticity $\eta$ shown for
$J^P=0^+$ and $1^+$ remains large, indicating that these channels are coupled
weakly. Figure \ref{poles} summarizes the findings, showing the complex poles
determined from the lattice data. For more details on the analysis please
refer to \cite{Dudek:2014qha,Wilson:2014cna}.

\begin{figure}[tbh]
\begin{center}
\includegraphics[width=0.45\textwidth,clip]{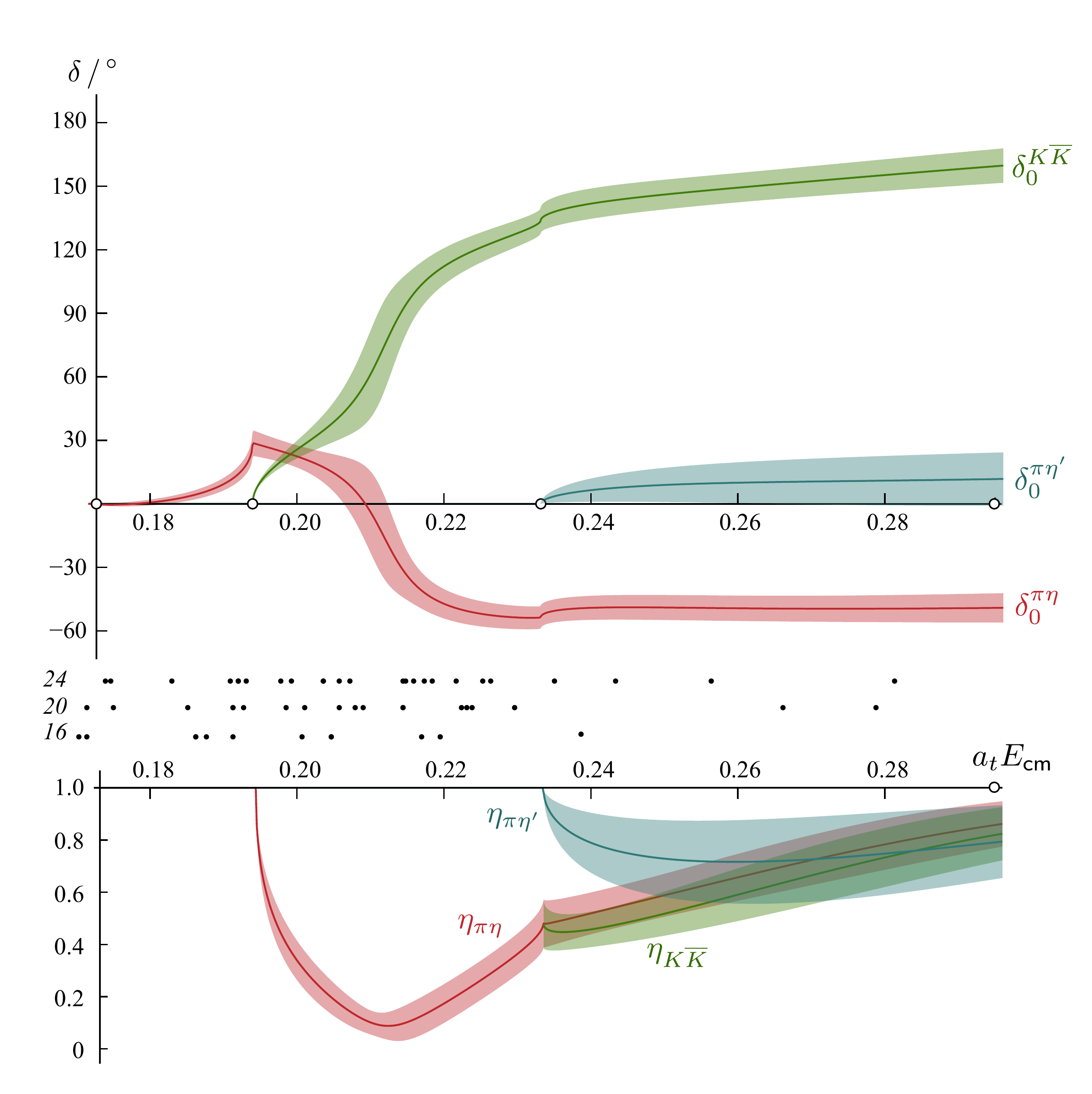}
\includegraphics[width=0.45\textwidth,clip]{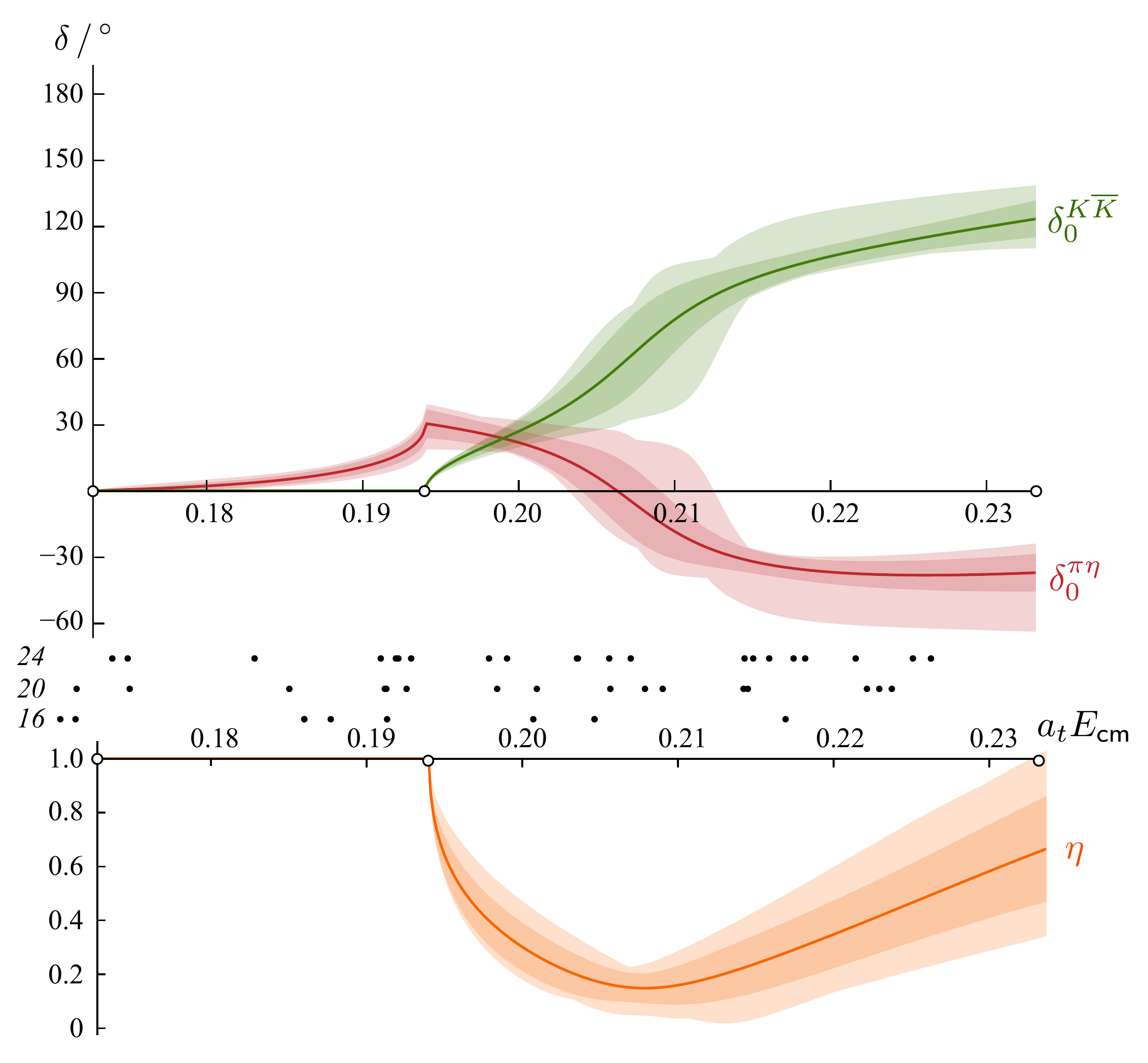}
\caption{\label{pieta_kbark} Left pane: The S-wave phase shift and inelasticities
  with three coupled channels ($\pi\eta$, $K\bar{K}$) and $\pi\eta^\prime$ in
the isovector, scalar ($I(J^P)=1(0^+)$) channel. Right pane: Both plots from \cite{Dudek:2016cru}.}
\end{center}
\end{figure}

An example for a calculation involving tightly coupled channels is given by $\pi\eta$--$K\bar{K}$
scattering \cite{Dudek:2016cru} with quantum numbers $I(J^P)=1(0^{+})$. Here up to three coupled
channels (also including $\pi\eta^\prime$) were considered. The lhs panel of Figure
\ref{pieta_kbark} shows the corresponding results for the phase shifts and
inelasticities at a pion mass of $391$~MeV. The rhs panel illustrates the
systematic uncertainties from varying the parameterizations used below the
$\pi\eta^\prime$ threshold. All successful parameterizations lead to a pole
close to the $K\bar{K}$ threshold on a single unphysical Riemann sheet. This
pole is likely related to the physical $a_0(980)$. For a full discussion of
the results the reader is referred to \cite{Dudek:2016cru}. In
addition the lattice data for the D-wave scattering features a narrow
resonance, possibly related to the physical $a_2(1320)$.

\subsection{A lesson from meson-meson scattering}
\label{lesson}

Studies of meson-meson scattering have demonstrated that a diverse basis of
lattice interpolating operators is needed to reliably extract the finite
volume spectrum. For hadron-hadron scattering this means in particular that to
ensure good overlap with the physical states, all relevant meson-meson or meson-baryon interpolators should be included
explicitly in the correlator basis.

\begin{figure}[tbh]
\begin{center}
\sidecaption
\includegraphics[clip,width=0.6\textwidth]{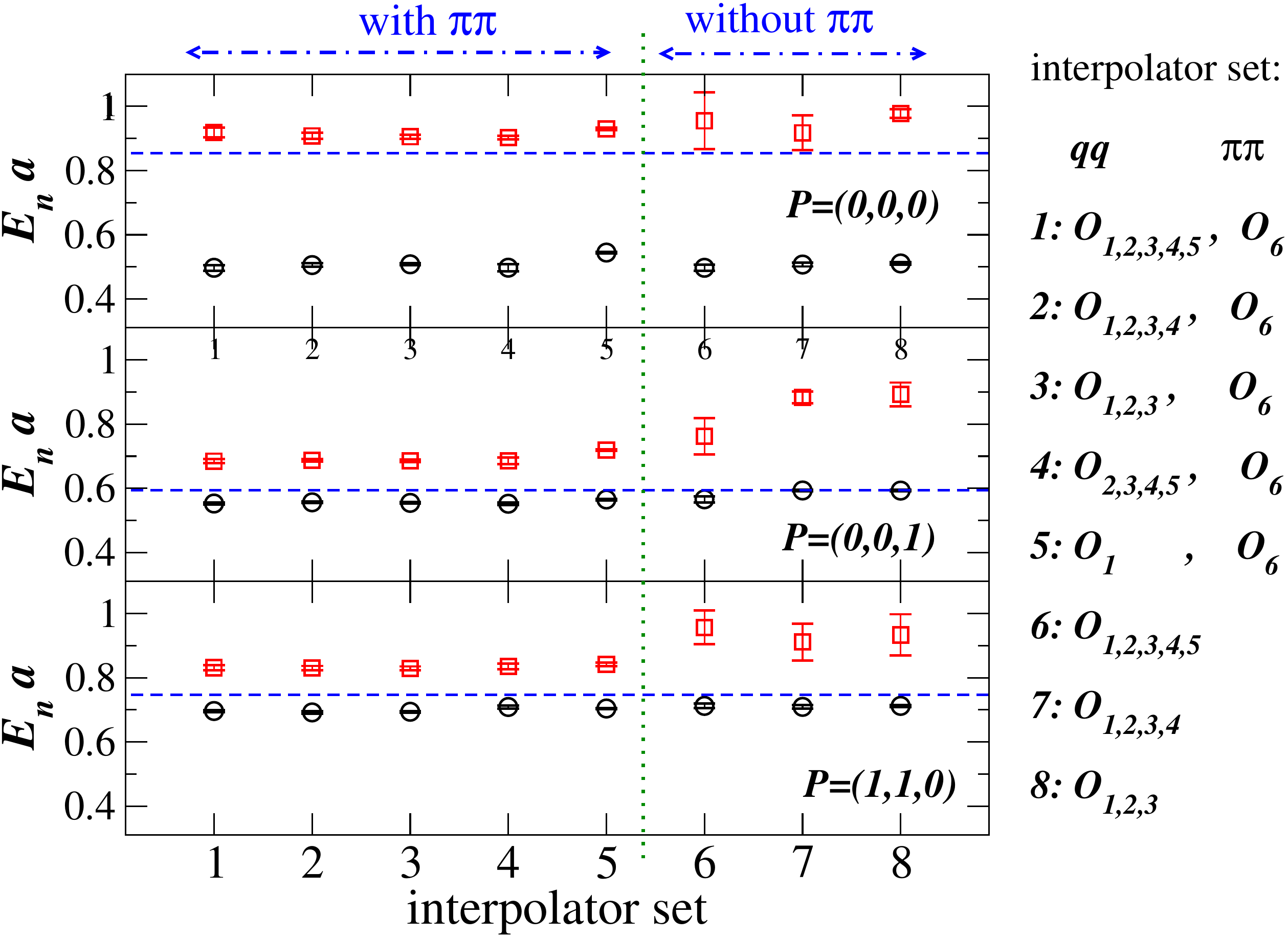}
\caption{\label{rho_mohler} The lowest two energy levels from isospin-1
  $\pi\pi$ scattering extracted using   different sub-matrices (interpolators
  sets) of the full $16\times16$ correlation matrix. For details on the
  basis used please refer to \cite{Lang:2011mn}.}
\end{center}
\end{figure}

\begin{figure}[tbh]
\begin{center}
\sidecaption
\includegraphics[clip,width=0.6\textwidth]{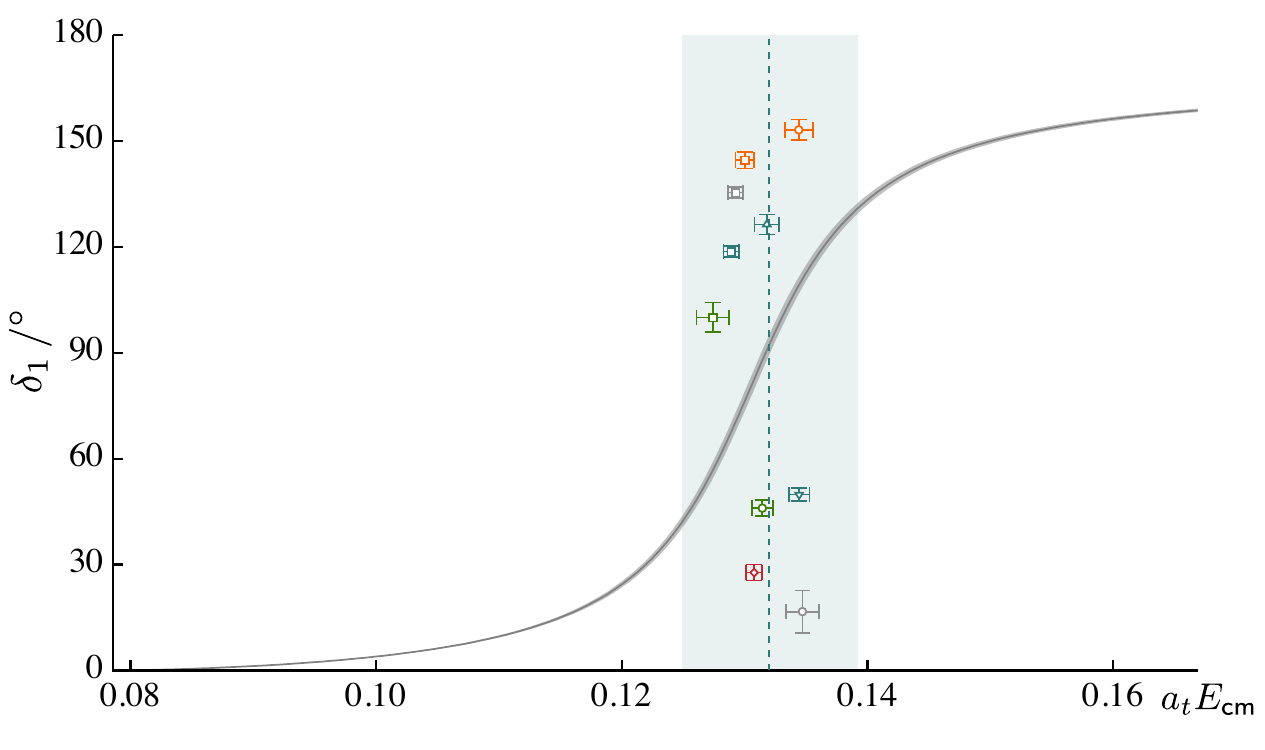}
\caption{\label{rho_jlab_1level} Elastic phase shift points extracted using
  only single-hadron-like interpolators (data points) compared to the
  Breit-Wigner parameterization of the phase shift points in Figure
  \ref{rho_jlab}. The vertical band indicates the Breit-Wigner mass and
  width. Figure from \cite{Wilson:2015dqa}.}
\end{center}
\end{figure}

To illustrate this point further it is instructive to take a look at how the
simulation results change when the full basis used in a given study is
artificially truncated. This is illustrated in Figures \ref{rho_mohler} and
\ref{rho_jlab_1level}. Figure \ref{rho_mohler} shows results for low-lying
energy in three momentum frames from an early simulation of $\pi\pi$
scattering in the $\rho$ meson channel \cite{Lang:2011mn}. When the basis consisting of
both quark-antiquark and meson-meson interpolating operators is reduced to
contain just quark-antiquark operators, some of the excited-state energy
levels become either ill-determined or display fake-plateaus, not
representative of the physical spectrum. Figure \ref{rho_jlab_1level} from
\cite{Wilson:2015dqa} provides an illustration of the same problem: Using only the energy levels
from single hadron (quark-antiquark) interpolators, the true phase-shift shown
with the fit extracted from the full data set (curve) is missed and one would
naively extract a much narrower resonance.

\subsection{Baryon resonances in Meson-Baryon scattering}
\label{baryons}

Studies of baryon-meson scattering on the lattice are complicated by several
aspects. For one, the signal to noise ratio at large Euclidean times is
exponentially suppressed, leading to much noisier data. Furthermore the number
of possible contractions is usually larger and the objects one calculates are
more complex. Frames with total nonzero momentum lead to parity
mixing in the lattice data, which makes the extraction of the spectrum even
more challenging.

Consequently most studies of baryon spectroscopy to date extracted the energy
levels from three-quark interpolators only. In most cases multi-hadron levels
are absent from the resulting spectra, and the cautionary remarks from Section
\ref{lesson} also apply to these studies. While showing some qualitative
features of the physical spectrum, it is unclear to what extent such studies
can be trusted. In this section, we therefore focus on exploratory studies
including baryon-meson or other five-quark interpolating fields.

\begin{figure}[tbh]
\begin{center}
\sidecaption
\includegraphics[clip,width=0.5\textwidth]{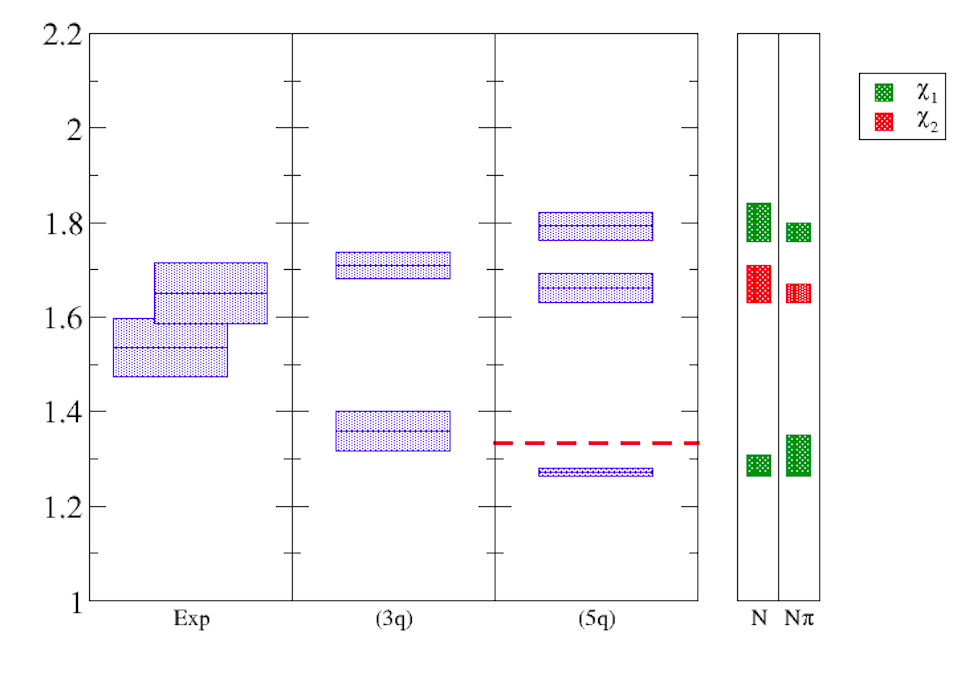}
\caption{\label{verduci_lang} Results for the S-wave negative parity Nucleon
spectrum. The left column shows the resonances observed in experiment shifted
up by $\Delta E=130$~MeV. The other columns show the lattice results from
3-quark interpolators alone (3q) and from the full basis consisting of 3-quark
and Nucleon-pion interpolators. For a full explanation see \cite{Verduci:2014csa}.}
\end{center}
\end{figure}

A first study of low-lying negative parity nucleons has been performed by Lang
and Verduci \cite{Lang:2012db,Verduci:2014csa}. Figure \ref{verduci_lang} shows their results on a
single volume with a pion mass of $266$~MeV. Just like the meson studies
discussed before, the spectrum using just single hadron interpolators is
incomplete. Once meson-baryon interpolators are included, the low-lying energy
spectrum can be extracted with reasonable precision. Future lattice QCD
studies of negative parity baryon resonances will however require more volumes and/or the use
of moving frames as well as additional scattering channels.

\begin{figure}[tbh]
\begin{center}
\sidecaption
\includegraphics[clip,width=0.6\textwidth]{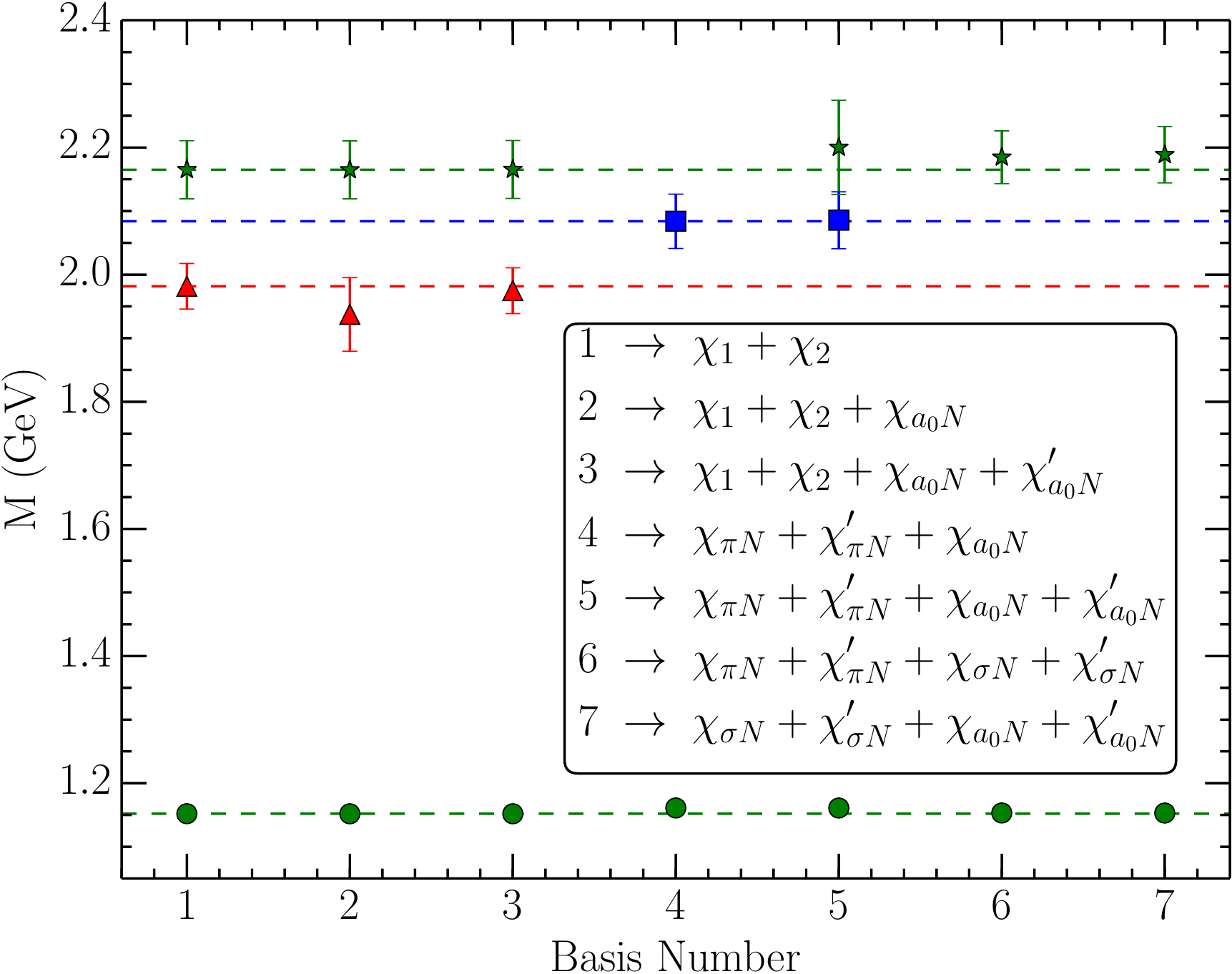}
\caption{\label{roper_adelaide} Low lying states from various sub-matrices of
  the full interpolator basis. $\chi_1$ and $\chi_2$ correspond to standard
  3-quark nucleon interpolator, while $\chi_{a_0N}$, $\chi_{\pi N}$, and
  $\chi_{\sigma N}$ are local 5-quark interpolators with the structure of
  these meson-baryon combinations. Figure from \cite{Kiratidis:2016hda}.}
\end{center}
\end{figure}

A more recent study by the Adelaide group \cite{Kiratidis:2016hda} explored the use of local
5-quark interpolators for extracting the positive parity nucleon
spectrum. Figure \ref{roper_adelaide} shows the dependence of the energy
levels observed when varying the interpolator basis. The complete basis
consisted of 3-quark and local 5-quark interpolators with structures
resembling $N\pi$, $N\sigma$, and $N a_0$ states. Unlike \cite{Lang:2012db} this study
does not result in energy levels close to the relevant thresholds and the
spectrum should probably considered to be incomplete. The authors suggest that
their results favor a scenario where the Roper resonance is a dynamically
generated resonance.

\section{Summary and outlook}

\label{outlook}

Recent years have seen a lot of activity with regard to exploratory studies of
resonances and bound states in meson-meson scattering. In the last two years
the first coupled-channel simulations have appeared. At the same time
calculations of meson-baryon energy levels are still scarce and a rigorous
extraction of low-lying baryon resonances from such calculations is facing
both conceptional and computational obstacles. Since the talk these
proceedings are based on, further results on the $\rho$ meson \cite{Fu:2016itp,Leskovec:2016lrm,Erben:2016zue} and on baryon-meson scattering in the Roper channel
\cite{Lang:2016hnn} have appeared. Furthermore there are a number of results from meson-meson
scattering involving heavy quarks which I could not cover. For a recent review of
those results please refer to \cite{Prelovsek:2015fra}.

All of these calculations are currently of an exploratory nature, and future
lattice studies will have to further address the systematic uncertainties, for example
arising from discretization effects, exponential volume corrections, the
pion mass dependence of observables, and, in some cases, neglected
three-particle channels. In light of the many challenges one should recall
that, unlike approaches based on models, the results of this programme are
parameter free predictions of QCD.

\bibliography{MOHLER_Daniel_CONF12}

\end{document}